\documentclass[aps,pra,twocolumn,superscriptaddress]{revtex4}

\usepackage{graphicx}
\usepackage{amssymb}
\usepackage{subfigure}
\usepackage{amsmath}
\usepackage{amsfonts}
\usepackage{bm}
\usepackage{color}

\usepackage{epsfig}

\newcommand{\ket}[1]{\vert#1\rangle}
\newcommand{\bra}[1]{\langle#1\vert}

\usepackage{graphicx}
\usepackage{bm}
\newcommand{\oper}[1]{\boldsymbol{\mathsf{#1}}}

\begin{document}

\title{Entanglement breaking channels and entanglement sudden death}


\author{Laura T. Knoll}
\affiliation{DEILAP, CITEDEF \& CONICET, J.B. de La Salle 4397, 1603 Villa Martelli, Buenos Aires, Argentina}
\author{Christian T. Schmiegelow}
\affiliation{Laboratorio de Iones y \'Atomos Frí\'{\i}os, Departamento de F\'{\i}sica, FCEN, UBA \& IFIBA, CONICET, Pabell\'on 1, Ciudad Universitaria, 1428 Buenos Aires, Argentina}
\author{Osvaldo~Jim\'enez~Far\'{\i}as}
\affiliation{ICFO-Institut de Ciencies Fotoniques, The Barcelona Institute of Science and
Technology, 08860 Castelldefels (Barcelona), Spain”}
\affiliation{Centro Brasileiro de Pesquisas F\'isicas, Rua Dr. Xavier Sigaud 150, Rio de Janeiro, 22290-180 Rio de Janeiro, Brazil}

\author{Stephen P. Walborn}
\affiliation{Instituto de F\'{\i}sica, Universidade Federal do Rio de
Janeiro, Caixa Postal 68528, Rio de Janeiro, RJ 21941-972, Brazil}

\author{Miguel A. Larotonda}
\affiliation{DEILAP, CITEDEF \& CONICET, J.B. de La Salle 4397, 1603 Villa Martelli, Buenos Aires, Argentina}

\date{\today}

\begin{abstract}

The occurrence of entanglement sudden death in the evolution of a bipartite system depends on both the initial state and the channel responsible for the evolution. An extreme case is that of entanglement braking channels, which are channels that acting on only one of the subsystems drives them to full disentanglement regardless of the initial state. In general, one can find certain combinations of initial states and channels acting on one or both subsystems that can result in entanglement sudden death or not. Neither the channel nor the initial state, but their combination, is responsible for this effect, but their combination. In this work we show that, in all cases, when entanglement sudden death occurs, the evolution can be mapped to that of an effective entanglement breaking channel on a modified initial state. Our results allow to anticipate which states will suffer entanglement sudden death or not for a given evolution. An experiment with polarization entangled photons demonstrates the utility of this result in a variety of cases. 
\end{abstract}

\pacs{}

\maketitle

\section{Introduction}

Quantum entanglement is a property of physical systems composed of two or more parts and is a consequence of the superposition principle on bipartite or multipartite systems. It takes the form of correlations of measurement results that cannot be reproduced by any classical mechanism. For this reason, entanglement has become a physical concept of central importance for the foundations of Quantum Mechanics \cite{schrodinger1935discussion}.  
Apart from its conceptual relevance, entanglement is also a resource that can be used to accomplish informational tasks like teleportation \cite{bennett1993teleporting}, quantum key distribution and super dense coding \cite{horodecki2009quantum,gisin2002quantum,agrawal2006perfect}.

The interaction of an entangled system with its environment results in an irreversible distribution of the entanglement between the system and the environment \cite{aguilar2014experimental,zurek2003decoherence,aguilar2014flow,dur2000three,xu2009experimental,aolitarev}. Interestingly, through the so-called Choi-Jamiolkowski relation \cite{jamiolkowski1972linear,choi1975completely}, the representation of a quantum channel and a state are equivalent within the theory.  In this report we explore this connection between environments and states and its implications for the dynamics of entanglement. We perform experiments where we produce pairs of polarization entangled photons, initially in a variety of entangled mixed states. Then, one of the photons is exposed to a local environment.  For a given environment $E$, some states of the system $S$ suffer Entanglement Sudden Death (ESD), while for other states the entanglement will disappear only asymptotically, as the interaction time tends to infinity \cite{almeida2007environment,laurat2007heralded,eberly2007end,yu2009sudden,aolitarev,drumond2009asymptotic,cunha2007geometry}. Theoretical considerations allow us to show that ESD occurs if and only if there exists a local effective Entanglement-Breaking Channel (EBC). This effect arises as the interplay between an initial state and a channel that annihilates any entanglement that one could try to establish through it \cite{horodecki2003entanglement,ruskai2003qubit}. The effective EBC is not the channel $E$ nor the state of the system $S$, rather it is a combination of the properties of both of them that we are able to measure experimentally. We show that this allows one to anticipate which states will suffer ESD or not for a given channel. This represents our main result. 

Section \ref{chanstate} is devoted to review and define notation on the duality between quantum channels and states, for a general $d$-dimension bipartite system. In section \ref{qubitevo} we address the problem of the entanglement evolution on a two-qubit system, and we explicitly show the relation between entanglement breaking channels and the sudden death of entanglement. Finally we apply these results in section \ref{exper} to two experimental situations using polarization-entangled photon pairs, where one of the qubits evolves through a noisy channel, by interacting with a controlled environment implemented using it's path internal degree of freedom. 

\section{The relation between channels and states}
\label{chanstate}
The evolution of a system $S$ of dimension $d$ due to an environment $E$, can be represented by at most $d^2$ Kraus operators, $\oper{K}_i$ by
\begin{eqnarray}
\oper{\mathcal{E}}[\rho_S]=\sum_{i=1}^{d^2}\oper{K}_i\rho_S\oper{K}_i^{\dagger},
\end{eqnarray}
where $\rho_S$ is the the state of the system. To conserve probabilities, we choose to work with trace preserving maps. Such assumption implies that the Kraus operators must satisfy
\begin{eqnarray}
\sum_{i=1}^{d^2}\oper{K}_i^{\dagger}\oper{K}_i=\oper{\mathcal{I}}.
\end{eqnarray}
Time dependence of the Kraus operators $\oper{K}_i(t)$ is omitted from this notation, for simplicity.

A mathematical relation can be established between channels and states; this is the aforementioned Jamiolkowski-Choi (J-CH) isomorphism \cite{jamiolkowski1972linear,choi1975completely}. More than just a theoretical tool the so-called \textit{Channel-State duality} has many practical implications and can be stated as follows: 

The set of channels $\{\oper{\mathcal{E}}\}$ acting on $\mathcal{C}^d$ is isomorphic to the set of bipartite states $\{\rho^* \}$  in $\mathcal{C}^d\otimes \mathcal{C}^d$, satisfying ${\rm Tr}_{2}[\rho^*]=\frac{1}{d}$, where $\mathcal{C}^d$ is a $d$-dimensional Hilbert space and ${\rm Tr}_{2}[\rho^*]$ is the partial trace over the second system \cite{horodecki1999general}.

\begin{figure}
\begin{center}
\includegraphics[width=8cm]{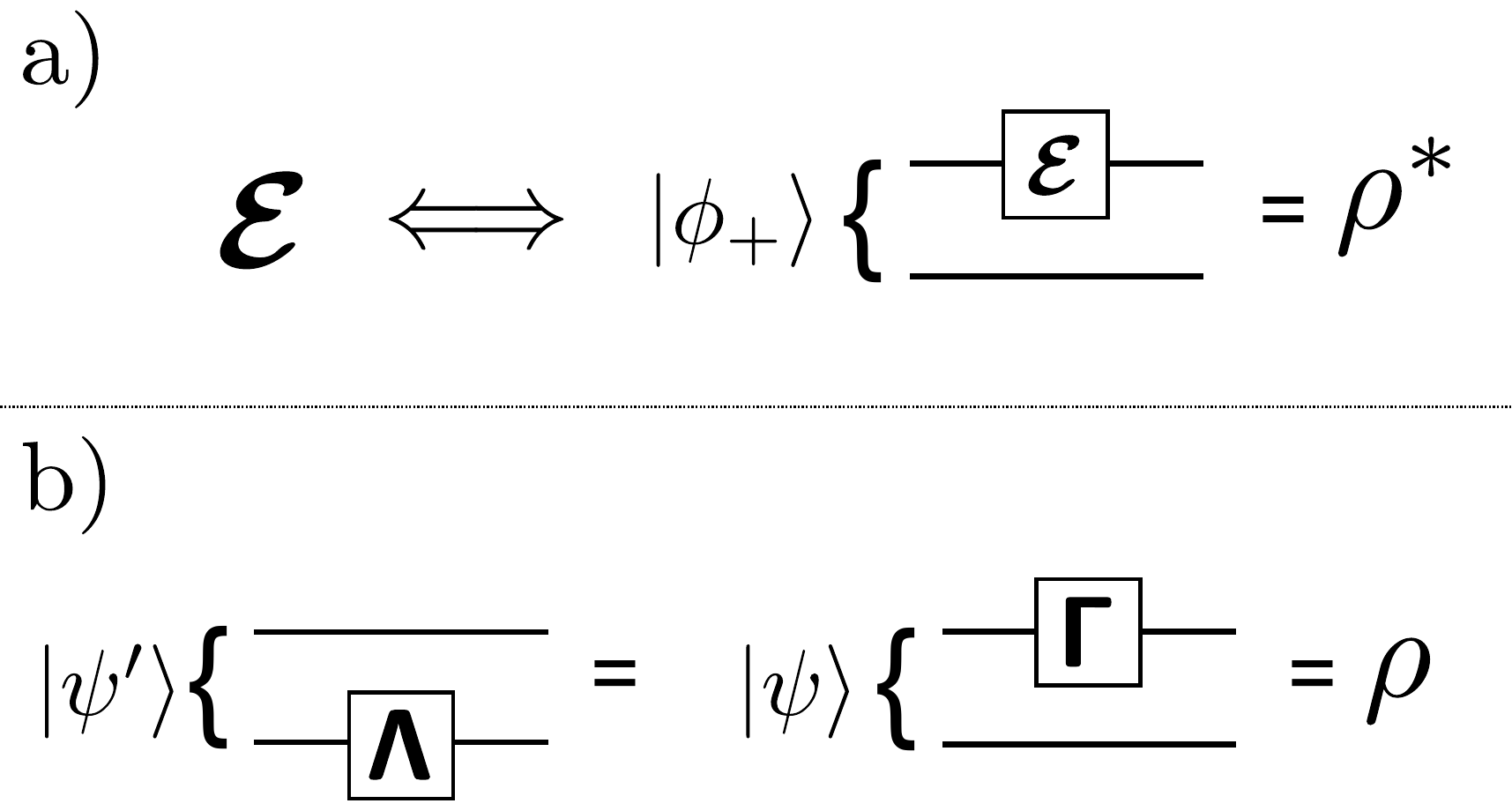}
\caption{\label{fig:chanstate} The relation between channels and states. a) Any trace preserving channel corresponds to a bipartite state. b) However an arbitrary bipartite state can correspond to two non-equivalent channels as ilustrated here by means of state preparations departing from two non-equivalent pure states.}
\end{center}
\end{figure}

The isomorphism can be established using a maximally entangled state $\ket{\phi_+}=\frac{1}{\sqrt{d}}\sum_i \ket{i}\ket{i}$ as illustrated in fig \ref{fig:chanstate}a). For two qubits one then has that $\rho^*=(\oper{\mathcal{I}}\otimes\oper{\mathcal{E}})\ket{\phi_+}\bra{\phi_+}$ satisfies the isomorphism. 

The relation between channels and states can be extended for arbitrary states in $\mathcal{C}^d\otimes\mathcal{C}^d$ beyond the J-CH isomorphism, removing the condition ${\rm Tr}_{2}[\rho^*]=\frac{1}{d}$: it was shown by Werner \cite{werner2001quantum} that there always exists a channel $\oper{\Gamma}$ and a pure state $\sigma$  such that any bipartite density matrix $\rho$ can be written as $\rho=(\oper{\mathcal{I}}\otimes\oper{\Gamma})\sigma$, where $\sigma$ is a pure-state density matrix. 

This result associates two physical objects to a mixed entangled state: A pure state $\sigma$,  and a channel $\oper{\Gamma}$. In contrast to the J-CH isomorphism where there is a one to one correspondence between the target state and a unilateral channel, here the combination of a channel and a pure state is not unique: an arbitrary mixed state $\rho$ can be written as:

\begin{eqnarray}
\label{decomp}
\rho= (\oper{\mathcal{I}}\otimes\oper{\Gamma})\sigma= (\oper{\Lambda}\otimes\oper{\mathcal{I}})\sigma'
\end{eqnarray}
with $\oper{\Gamma} \neq \oper{\Lambda}$, and $\sigma  \neq \sigma'$ are two distinct pure states. This represents two ways of preparing the bipartite state $\rho$ as illustrated in Fig \ref{fig:chanstate}b).

\section{Qubit to qubit entanglement and its evolution}
\label{qubitevo}

 Consider a quantum operation $\oper{\mathcal{E}}$ acting in one of the qubits, so that the bipartite state evolves through $\oper{\mathcal{I}}\otimes \oper{\mathcal{E}} $. Konrad \emph{et al.}~\cite{konrad2008evolution} demonstrated that the concurrence~\cite{wootters2001entanglement} of pure bipartite states, when exposed to one qubit channels, evolves according to 
\begin{eqnarray}
C[(\oper{\mathcal{I}}\otimes\oper{\mathcal{E}})\ket{\chi}\bra{\chi}]= C[(\oper{\mathcal{I}}\otimes\oper{\mathcal{E}})\ket{\phi_+}\bra{\phi_+}]C[\ket{\chi}]. 
\label{factpuro}
\end{eqnarray}

This equation shows that, for pure states, the entanglement evolution depends on the state only by means of the initial concurrence $C[\ket{\chi}]$; all the information about the evolution of the entanglement is contained in the dual state of the channel $\oper{\mathcal{E}}$.

We can also derive an evolution equation for the concurrence of mixed states. According to Eq.~\eqref{decomp} one can consider a mixed state as the combination of a pure state and a unilateral channel. Then, under the action of the channel $\oper{\mathcal{E}} $, a mixed state $\rho$ can be written as     $C\left[(\oper{\mathcal{I}}\otimes\oper{\mathcal{E}})\rho\right]=C\left[(\oper{\mathcal{I}}\otimes\oper{\mathcal{E}})(\oper{\mathcal{I}}\otimes\oper{\Gamma})\sigma\right]$, where $\sigma$ a pure state. It follows from Eq. \eqref{factpuro} that
\begin{equation}
\label{general}
C\left[(\oper{\mathcal{I}}\otimes\oper{\mathcal{E}})\rho\right]=C\left[(\oper{\mathcal{I}}\otimes\oper{\mathcal{E}}\oper{\Gamma})|\phi_+\rangle\langle\phi_+|\right]C[\sigma].
\end{equation}

This is the extended version of Eq.~\eqref{factpuro}, which was reported and experimentally tested in Ref. \cite{farias2009determining}. It shows that, even if the input state is mixed, the final concurrence is given by the product of two factors: the concurrence of a Bell state evolving under the action of the product of two channels $\oper{\mathcal{E}}\oper{\Gamma}$, acting on the second qubit, and the concurrence of a pure state, $C[\sigma]$. Both the channel $\oper{\Gamma}$ and the factor $C[\sigma]$ depend only on the initial mixed state.  
\par

With these tools, we now study the situation where entanglement in the qubits vanishes in finite time, a phenomenon better known as Entanglement Sudden Death (ESD).
\subsection{Unilateral Channels - Pure States}
 Let's consider first the situation of a pure initial state $\ket{\chi}$ subject to a unilateral channel, $\oper{\mathcal{I}}\otimes\oper{\mathcal{E}}$. In this case the evolution equation \eqref{factpuro} says that the concurrence depends, up to a constant factor, only on the channel through $C[(\oper{\mathcal{I}}\otimes\oper{\mathcal{E}})\ket{\phi_+}\bra{\phi_+}]$. No matter what the initial entangled pure state is, if the entanglement vanishes there is only one cause for this: the channel $\oper{\mathcal{E}}$. This is the definition of an Entanglement Breaking Channel (EBC) rephrased in terms of the evolution equation. Many channels such as Amplitude Damping (ADC) or Phase Damping (PDC) are EBCs only asymptotically, and so they do not produce ESD alone. An example of an EBC for qubits is the Depolarizing channel \cite{nielsen2010quantum}. 
\subsection{Unilateral Channels - Mixed States}
Now we consider a \emph{mixed} entangled state $\rho$ evolving under the action of $\oper{\mathcal{I}}\otimes \oper{\mathcal{E}} $. In this situation the evolution depends on the initial state $\rho$, and Eq.\eqref{general} tells us exactly how. If the channel $\oper{\mathcal{E}}$ is an EBC for finite times, then it is clear that the composition will be also an EBC and ESD will occur. On the other hand, if the channel $\oper{\mathcal{E}}$ is not an EBC, ESD may still occur because of the mixture of the state $\rho$.  In both cases there is an effective EBC $\oper{\mathcal{E}} \oper{\Gamma}$ acting on a maximally entangled state, that is responsible for the ESD. Indeed, an initially mixed state defines a whole family of EBC's through the equation
\begin{equation}
C[(\oper{\mathcal{I}}\otimes\oper{\mathcal{E}}\oper{\Gamma})\ket{\phi_+}\bra{\phi_+}]=0.
\end{equation}
Here we see the usefulness of the generalized evolution equation, as it separates correctly the characteristics of the mixed states that lead to a decrease or loss of entanglement.
  
\subsection{Bilateral Channels - Pure and Mixed States}
Consider now two different pure bipartite entangled states $\ket{\Phi_1}$ and $\ket{\Phi_2}$ under the action of two channels $ \mathcal{E}_1$ and $\mathcal{E}_2$, such that none of these channels produce ESD by itself; i.e.

\begin{align*}
C[(\oper{\mathcal{E}}_1\otimes\oper{\mathcal{I}})\ket{\Phi_i}\bra{\Phi_i}]&\neq0,\\ C[(\oper{\mathcal{I}}\otimes\oper{\mathcal{E}}_2)\ket{\Phi_i}\bra{\Phi_i}]&\neq0,
\end{align*}
$i=1,2$. We may consider the action of both channels acting on $\ket{\Phi_1}$ and $\ket{\Phi_2}$, such that
\begin{eqnarray}
C[(\oper{\mathcal{E}}_1\otimes\oper{\mathcal{E}}_2)\ket{\Phi_1}\bra{\Phi_1}]=0,
\label{sudden2}
\end{eqnarray}
while
\begin{eqnarray}
C[(\oper{\mathcal{E}}_1\otimes\oper{\mathcal{E}}_2)\ket{\Phi_2}\bra{\Phi_2}]\neq0. \label{sudden}
\end{eqnarray}

This condition was first observed in \cite{almeida2007environment}. In view of Eq. (\ref{sudden2}) we notice that by use of Eq. (\ref{decomp}) one can find a channel $\oper{\Gamma}$ and a pure  state $\Sigma$ such that  
\begin{align*}
 (\oper{\mathcal{E}}_1\otimes\oper{\mathcal{E}}_2)\ket{\Phi_1}\bra{\Phi_1}&= (\oper{\mathcal{I}}\otimes\oper{\mathcal{E}}_2)(\oper{\mathcal{E}}_1\otimes\oper{\mathcal{I}})\ket{\Phi_1}\bra{\Phi_1}\\
&= (\oper{\mathcal{I}}\otimes\oper{\mathcal{E}}_2)(\oper{\mathcal{I}}\otimes\oper{\mathcal{E}'}_1)\Sigma\\
&= (\oper{\mathcal{I}}\otimes\oper{\Gamma})\Sigma.
\end{align*}
By applying Eq. (\ref{factpuro}), the concurrence depends on the action of $\oper{\Gamma}$ on a maximally entangled state and on the concurrence of the state $\Sigma$; satisfying Eq. (\ref{sudden2}) then implies
\begin{eqnarray}
C[(\oper{\mathcal{I}}\otimes\oper{\Gamma})\Sigma]= C[(\oper{\mathcal{I}}\otimes\oper{\Gamma})\ket{\phi_+}\bra{\phi_+}]C[\Sigma]=0.
\label{macelo1}
\end{eqnarray} 
Since $\oper{\mathcal{E}}_1$ does not produce ESD by itself as stated above, we obtain
\begin{align*}
0 &\neq C[(\oper{\mathcal{E}}_1\otimes\oper{\mathcal{I}})\ket{\Phi_1}\bra{\Phi_1}]=C[(\oper{\mathcal{I}}\otimes\oper{\mathcal{E}'}_1)\ket{\phi_+}\bra{\phi_+}]C[\Sigma],
\end{align*}
and therefore we know that $C[\Sigma]\neq 0$. Hence, according to Eq. (\ref{macelo1}), ESD is caused by the action of the effective channel $\oper{\Gamma}$ on a maximally entangled state, which is indeed an EBC. 

In the same way, the channel-state combination of Eq. (\ref{sudden}) can be written as a channel $\oper{\Lambda}$ acting on another pure state $\Sigma'$:
\begin{eqnarray}
 (\oper{\mathcal{E}}_1\otimes\oper{\mathcal{E}}_2)\ket{\Phi_2}\bra{\Phi_2}=(\oper{\mathcal{I}}\otimes \oper{\Lambda})\Sigma'
 \end{eqnarray} 
where, by following the same steps that lead to Eq. (\ref{macelo1}) we find that $\oper{\Lambda}$ is not an EBC.

In summary, this shows our claim that for every Entanglement Sudden Death, one can associate an effective Entanglement Breaking Channel acting on only one of the subsystems.

\section{Experimental Investigation of Entanglement Dynamics}
\label{exper}

\begin{figure}
\begin{center}
\includegraphics[width=8cm]{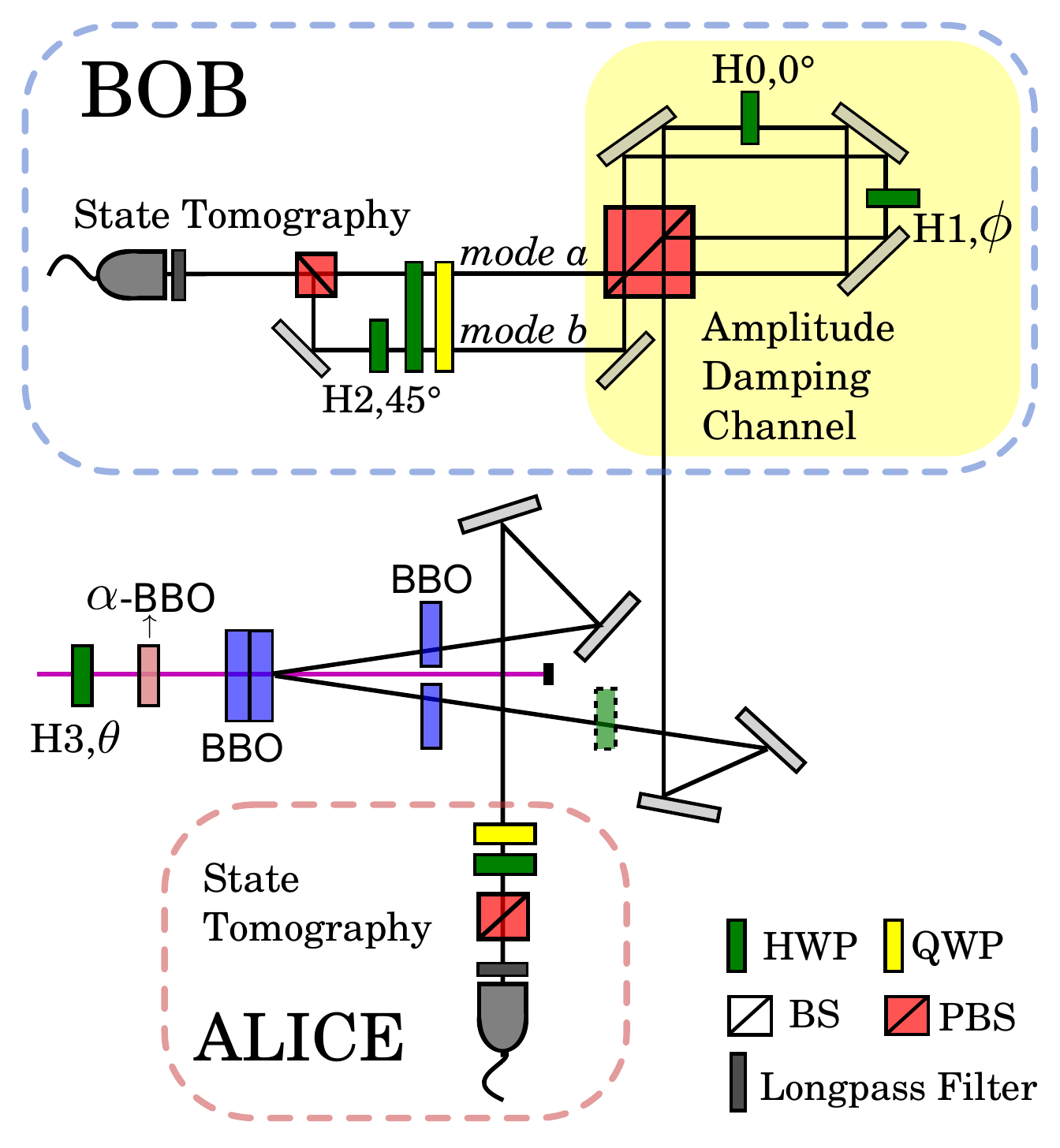}
\caption{\label{fig:setup}Experimental setup. The entangled SPDC pair source is optimized using temporal and spatial compensating birefringent crystals. Bob's polarization qubit can be coupled to a local environment encoded on the path qubit (see text). The light green-dashed HWP placed on Bob's path is occasionally used to generate state $\rho_2$ to simulate the noisy channel on Alice's side.  }
\end{center}
\end{figure}

In this section we describe an experiment where we observe the evolution of the entanglement of a pair of polarization qubits in two different scenarios and prove the use of the new interpretation presented in the previous sections.

The experimental set-up for the investigation of the entanglement dynamics is sketched in Fig. \ref{fig:setup}. Polarization entangled photon pairs are produced via spontaneous parametric down-conversion (SPDC) using a CW diode laser to pump a nonlinear crystal arrangement. A more detailed description of the entangled pair source can be found in \cite{knoll2014remote,knoll2014noisy}. After implementation of the specific quantum channels -which are described below- the photons from both paths, after passing through specific quantum gates implemented for the experiments, are directed to polarization analysis schemes and then detected in coincidence with single-photon detectors, with a temporal coincidence window of 8ns. Full tomography of the two photon polarizations state was implemented.

With this setup we are able to prepare the following quantum state
\begin{align}
\ket{\alpha}= \alpha\ket{HH}+\sqrt{1-|\alpha|^2}\ket{VV}.
\label{eq:spdc}
\end{align}
The real amplitude $\alpha$ can be controlled by rotating the angle $\theta$ of a half-wave plate (H3 in figure \ref{fig:setup}) in the pump laser beam~\cite{kwiat1999ultrabright}. 
Rotating H3 to $45^{\circ}$ produces the state $\ket{HH}\bra{HH}$, and adding another HWP at $45^{\circ}$ on Bob's path can further transform the downconverted pair into the state $\ket{HV}\bra{HV}$.
Starting from state \eqref{eq:spdc} we were able to produce several families of initially mixed states by weighted averages of different pure states. In the next section we explain in detail the families of mixed entangled states that were created.

Open system dynamics can be induced on one of the polarization-encoded qubits by using its path internal degree of freedom. The polarization qubit is coupled to the path, or linear momentum qubit in a controlled manner using a displaced Sagnac interferometer, and the information on this path qubit is traced out before the polarization analysis. This interferometer has been shown to implement a few characteristic quantum channels, including the amplitude damping channel and the phase damping channel~\cite{salles2008experimental}:
Bob's photon enters a Sagnac interferometer based on a polarization beamsplitter (PBS), where the $H$ and $V$ polarization components are routed in different directions. If $\phi=0^{\circ}$ both polarization components are coherently recombined in the PBS and exit the interferometer in mode $a$. For any other angle of $\phi$ the $V$ component is transformed into an $H$ polarized photon with probability $g=\sin^{2}(2\phi)$ and exits the interferometer in mode $b$. Both modes are later incoherently recombined using a half waveplate (HWP) oriented at $45^{\circ}$ on mode $b$ (H2) and another polarizing beam splitter, which corresponds to a partial tracing operation over the environment.

\subsection{Amplitude damping channel acting on a mixed state}

\begin{figure}
\begin{center}
\includegraphics[width=8cm]{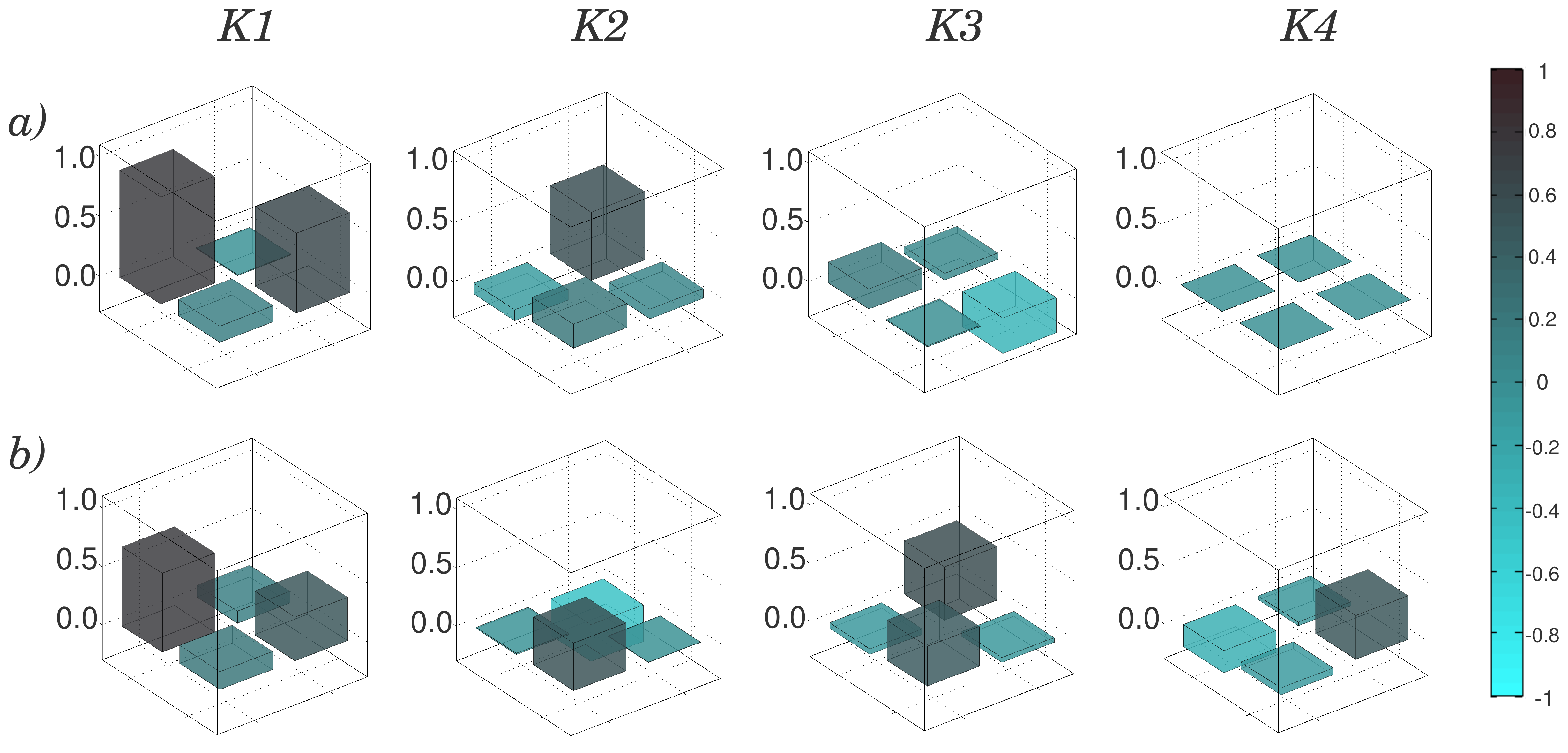}
\caption{\label{fig:kraus}Experimental Kraus operators: a) (top row) for the Amplitude Damping channel $\oper{\mathcal{E}}_g$ acting on one qubit; b) (bottom row) for the measured effective channel $\oper{\mathcal{E}}_g\oper{\Gamma}$, acting on an initial state with $\omega=0.25$ and $h=0.5$. In both cases the damping parameter is $g=0.6$.}
\end{center}
\end{figure}

We use quantum process tomography to characterize the effect of the Sagnac interferometer, set to implement an amplitude damping channel of strength $g$. By doing so we obtain the operator representation of the noisy channel $\oper{\mathcal{E}}_g$. The reconstructed Kraus operators are shown on Fig \ref{fig:kraus}a), for a damping parameter $g=0.6$. We want to observe the evolution of mixed states through the channel. 
We are able to prepare a family of mixed entangled states characterized by the parameter $p$,
\begin{equation}
 \rho=(1-p)\rho_1+p\rho_2\\
\label{simuA}
\end{equation}
where $\rho_{1}=\ket{\alpha}\bra{\alpha}$ given by \eqref{eq:spdc} 
 
and $\rho_2=\ket{HV}\bra{HV}$.

In this way, the noise on Alice's qubit is simulated as a weighted average of different experimentally obtained pure states, as described in \cite{knoll2014noisy}.

Once again, by using Eq. (\ref{decomp}), we can find a channel $\mathcal{E}_1$ acting on a pure state $\ket{\chi}$, such that
\begin{center}
 $\rho=(\oper{\mathcal{E}}_1\otimes\oper{\mathcal{I}})\ket{\chi}\bra{\chi}$.
\end{center}
 $\mathcal{E}_1$ may describe the situation where an amplitude damping channel with damping parameter $h$ acts on the first  qubit of the pure state $\ket{\chi}=\sqrt{\omega}\ket{HH}+\sqrt{1-\omega}\ket{VV}$.  In this way, $\rho$ is now characterized by the two parameters $h$ and $\omega$, such that $p=h(1-\omega)$.


Next we start the evolution through the amplitude damping channel (ADC) $\oper{\mathcal{I}} \otimes \oper{\mathcal{E}}_g$ on Bob's side and calculate the concurrence of the output state $\rho_{out}=(\oper{\mathcal{I}}\otimes \oper{\mathcal{E}}_g)\rho=(\oper{\mathcal{I}}\otimes\oper{\mathcal{E}}_g\oper{\Gamma})\sigma$. From the state tomography of initial state $\rho$, Eq.~\eqref{simuA} we can extract the map $\oper{\Gamma}$. Process tomography allows to obtain the Kraus operators corresponding to the ADC and combining those two, the effective channel $\oper{\mathcal{E}}_g\oper{\Gamma}$. A convenient representation of this channel is expressed through the Kraus operators

 \begin{equation*}
K_1=\begin{pmatrix}
\sqrt{\frac{\omega+gh(1-\omega)}{\omega+h(1-\omega)}} & 0\\
0 & \sqrt{\frac{\omega(1-g)}{\omega+gh(1-\omega)}}
\end{pmatrix}\; \; \;
K_3=\begin{pmatrix}
0 & \sqrt{g}\\
 0 & 0
\end{pmatrix}
\end{equation*}

\begin{equation}
K_2=\begin{pmatrix}
0 & 0\\
 \sqrt{\frac{h(1-g)(1-\omega)}{\omega+h(1-\omega)}} & 0
\end{pmatrix}\;\;\;
K_4=\begin{pmatrix}
0 & 0\\
 0 &  \sqrt{\frac{(1-g)hg(1-\omega)}{\omega+gh(1-\omega)}} 
\end{pmatrix}.
\label{adcadckraus}
\end{equation}

Figure \ref{fig:kraus}b) shows the measured Kraus operators for the effective channel, for the particular choice $\omega=0.25$ and $h=0.5$. These operators are measured for a damping parameter $g=0.6$.

 The concurrence of the family of initial states studied is 
\begin{equation}
C(\sigma)=2\sqrt{(1-h)(1-\omega)[\omega+h(1-\omega)]},
 \label{cdesigma}
\end{equation}
which is a positive function in the $(0,1)\times(0,1)$ interval. Therefore, the vanishing points of $C[(\oper{\mathcal{I}}\otimes\oper{\mathcal{E}}_g)\rho]$ are also the vanishing points of the concurrence of the map applied to a maximally entangled state, $C[(\oper{\mathcal{I}}\otimes\oper{\mathcal{E}}_g\oper{\Gamma})|\phi_+\rangle\langle\phi_+|]$. Entanglement breaking characteristics of the map can be studied with this scheme. With this in mind, we can monitor the evolution of the concurrence in the parameter space, where we remind that $g$ parametrizes the damping parameter of the noisy channel. 

 \begin{figure*}
  \includegraphics[width=\textwidth,height=5.5cm]{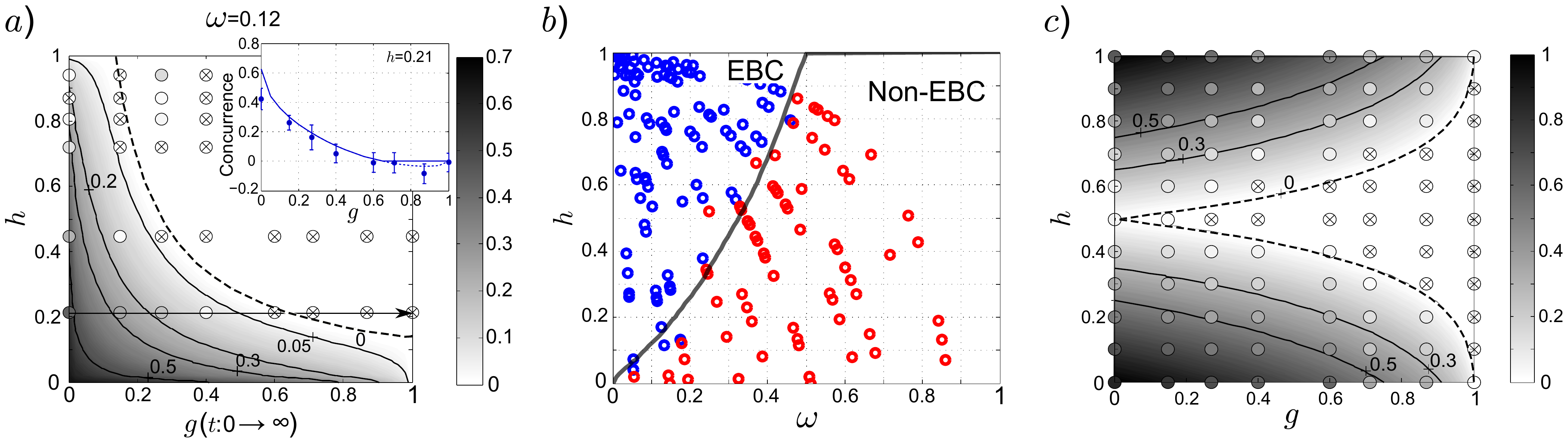}
  \caption{\label{fig:maps} Experimental results and theoretical predictions. a)  Evolution of the concurrence of states initially characterized by $\omega=0.12$, and different degrees of mixture $h$, under the interaction with a local ADC channel. Concurrence values are coded on the grayscale bar: the continuous map is the solution of Eq. \eqref{general}, while the symbols plot the measured concurrence for different experimental initial conditions and evolution parameter $g$. Inset: measured and theoretical concurrence evolution of a particular initial state ($h=0.21$) showing ESD. b) parameter regions that define the quantum map \eqref{adcadckraus} as an EBC or Non-EBC: any pair of values $h$, $\omega$ that fall into the upper left corner generate an EBC. The blue circles are initial conditions for which we
measured concurrence values of zero for some value of $g<$1, while the solid black line represents the theoretical boundary between EB and non-EB channels. c) Evolution of the concurrence of X-states characterized by their initial mixture $h$ under the action of local phase damping channels.  Contour lines are plotted in a) and c) as reference for certain concurrence values, while crossed-out circles indicate zero-concurrence experiment outcomes.}    
\end{figure*}


Figure \ref{fig:maps}a) shows gray-scale maps representing the concurrence values as a function of the degree of mixture $h$ and the noise parameter on Bob's qubit $g$, for states with initial degree of coherent superposition $\omega$=0.12. The colorbar on the right indicates the value of the concurrence. Contour lines are also plotted as a reference. The color within the circles represent the actual measured values of the concurrence. A symbol and its background with similar gray level means a good agreement between measurements and theory.

In figures \ref{fig:maps}a) and c), horizontal lines correspond to the evolution of the concurrence of a given initial state. 
As an example, the entanglement dynamics for $h=0.21$ and $\omega=0.12$ is represented in the inset at the top of figure \ref{fig:maps}a), where we can see that ESD occurs for $g\approx0.65$. The contour line for $\mathcal{C}=0$ shows whether the map applied to a specific initial state generates ESD or not, at some instance in the evolution through the damping channel. For $\omega\sim0.5$ there is no sudden death, as the initial input state becomes the maximally entangled state $\ket{\Phi_+}$ \cite{almeida2007environment}.

Thanks to Eq. \eqref{general}, we can find regions in the parameter space $\omega, h$ (initial coherent superposition and degree of mixture) for which the channel is EBC, i.e. where the concurrence vanishes for some value of $g<1$. This is depicted in figure \ref{fig:maps}b); for values of $\omega<$ 0.5, a certain initial value of $h$ converts the channel from non-EB to EB. From the quantum state point of view, figure \ref{fig:maps}b) simply shows which are the initial mixed states \eqref{simuA} that suffer ESD under the action of an AD channel. More interestingly, from the quantum channel point of view, this figure shows the entanglement-breaking capacity of the map $(\oper{\mathcal{I}}\otimes\oper{\mathcal{E}}_g\oper{\Gamma})$ specified in \eqref{adcadckraus}, with decay probability $g$: maps with parameters $h,\omega$ that falls on the left side of the boundary curve are EBC's, i.e. the concurrence vanishes for a finite evolution time ($g<1$), while maps with parameters that lie  on the right side of the 
figure 
are non-EBC. This feature was also experimentally verified: states with different values of $\omega,h$ were prepared, and the concurrence for increasing values of the interaction parameter $g$ was obtained by performing quantum state tomography. Initial conditions that lead to an experimental observation of the sudden death of entanglement are depicted with blue symbols, while conditions in which ESD was not observed for any evolution time are marked in red, showing a very good agreement with the theoretical prediction.

\subsection{Phase Damping Channel acting on X-states}
In a second set of measurements, we prepare the family of pure states $\rho_X=h\ket{\phi_+}\bra{\phi_+}+(1-h)\ket{\psi_+}\bra{\psi_+}$; both $\ket{\phi_+}$ and $\ket{\psi_+}$ are accessible experimentally. We can generate the state $\rho_X$ by averaging the results of the experiments using these two Bell states with their respective statistical weights, and calculate the concurrence for different values of $h$ and noise parameters. Again, $h$ characterizes the initial degree of mixture of the state. 

We study the dynamics by letting the state $\rho_X$ evolve through the phase damping channel (PDC), $\oper{\mathcal{E}}$  on Bob's side, obtaining $\rho^{PDC}=(\oper{\mathcal{I}}\otimes\oper{\mathcal{E}}'_g)\rho_X$. 
We can once again find a channel $\oper{\Lambda}$ and a pure state $\sigma^\prime$ such that $\rho_X=(\oper{\mathcal{I}}\otimes\oper{\Lambda})\sigma^\prime$. In this case, the channel $\oper{\Lambda}$ corresponds to a bit-flip channel with probability $h$, and $\sigma^\prime=\ket{\phi_+}\bra{\phi_+}$. 
Recalling \eqref{general}, we see that the concurrence of the output state, is just the concurrence of a Bell state evolved through the channel $\oper{\mathcal{E}}'_g\oper{\Lambda}_{h}$, since the concurrence of a Bell state $C[\ket{\phi_+}\bra{\phi_+}]=1$. 
Following the evolution of an X-state through local phase damping channels can therefore give information about the entanglement breaking capacity of the channel $\oper{\mathcal{E}}'_g\oper{\Lambda}_{h}$.
As in the previous experiment, the noise on Bob's side is implemented through the interaction with a path qubit in a polarization sensitive Sagnac interferometer, with a slight modification that allows for the implementation of a phase decay map \cite{almeida2007environment}.

Figure \ref{fig:maps}c) shows a gray-scale map representing the concurrence values 
of the Bell state $\ket{\phi_+}$ evolved through the channel $\oper{\mathcal{E}}'_g\oper{\Lambda}_{h}$, 
as a function of both $h$ and $g$. The measured values of the concurrence for the output state $\rho^{PDC}$ are plotted on top, with the same gray-scale coded values as the theoretical map.
X-states with $h$ close to 1/2 have a large degree of mixture. Evolution of the bit-flip channel can be observed by following the concurrence through vertical lines, while temporal evolution of the phase damping channel is obtained by tracing horizontal lines in the figure.
Accordingly, the concurrence of states with $h=1/2$ is 0 for all values of $g$.
On the other hand, for $h=0,1$ we obtain the states $\ket{\psi_+}, \ket{\phi_+}$ respectively, for which the concurrence drops to zero only when $g=1$. The complete map is EBC except for these limits.
As opposed to the ADC map, the phase damping has a symmetric behavior on the initial state populations. As the noise increases, the zero concurrence region becomes larger symmetrically with respect to $h$, due to the fact that there is no change in the populations.

\section{Conclusion}

We have presented experimental results for the dynamics of entanglement for mixed states, monitoring the evolution of the concurrence through noisy environments given by the amplitude damping and phase damping channels acting on different initial two-qubit states. Using the connection between channels and bipartite states, we were able to express the concurrence of the output state as the product of the concurrence of a Bell state evolved through an effective channel acting on a single qubit, and the concurrence of a pure state. In doing so, we could study the entanglement-breaking capacity of different effective channels, and to establish conditions on the map parameters that produce an EBC. We experimentally tested these conclusions by observing the evolution of an entangled state through different local damping channels. In particular, we studied the action of two local amplitude damping channels on a pure state and we related its dynamics to the evolution of a particular family of mixed states through an 
amplitude damping channel. The state-channel duality was also used to study the action of phase damping channels on X-states. An \emph{a priori} knowledge of the initial mixed state and/or the quantum channel that will affect the entangled resource could allow one to choose the optimal set of parameters for efficient quantum information processing.

\begin{acknowledgments}
 We acknowledge financial support from the Brazilian funding
agencies CNPq, CAPES, and FAPERJ, and the Argentine funding agencies CONICET and ANPCyT. This work was performed as part of the Brazilian National Institute of Science and Technology for Quantum Information. O.J.F. was supported by the Beatriu de Pin\'os fellowship (nº 2014 BP-B 0219) and Spanish MINECO (Severo Ochoa grant SEV-2015-0522). We thank Corey O'Meara for useful discussions. 
\end{acknowledgments}


%

\end{document}